Using Clinical Experts' Beliefs to Compare Survival Models in Health Technology Assessment

14 September 2021

Authors:

John W Stevens PhD

Martin Orr MSc

Corresponding author:

John W Stevens PhD (JohnWmStevens@gmail.com)

**Financial Disclosure**: None reported

**Acknowledgements**:

The authors would like to thank Professor Martyn Plummer for describing the concept of a soft constraint used to implement the zeros trick in JAGS, Professor Edward George for helpful discussions clarifying the implementation of dilution prior distributions, and Professor Anthony O'Hagan for insightful and constructive feedback on a draft version of this manuscript.

Any errors or other aspects not appreciated by reviewers are solely the responsibility of Dr Stevens.

**Abstract**

*Objectives:* The aim of this paper is to contrast the retrospective and prospective use of experts' beliefs in choosing between survival models in economic evaluations.

*Methods:* The use of experts' retrospective (posterior) beliefs is discussed. A process for prospectively quantifying prior beliefs about model parameters in five standard models is described. Statistical criterion for comparing models, and the interpretation and computation of model probabilities is discussed. A case study is provided.

*Results*: Experts have little difficulty in expressing their posterior beliefs. Information criterion is an approximation to Bayesian model evidence and is based on data alone. In contrast, Bayes factors measure evidence in the data and makes use of prior information. When model averaging is of interest, there is no unique way to specify prior ignorance about model probabilities. Formulating and interpreting weights of similar models should acknowledge the dilution phenomenon such that highly correlated models are given smaller weights than those with low correlation.

*Conclusion:* The retrospective use of experts' beliefs to validate a model is potentially misleading, may not achieve its intended objective and is an inefficient use of information. Experts' beliefs should be elicited prospectively as probability distributions to strengthen inferences, facilitate the choice of model, and mitigate the impact of dilution on model probabilities in situations when model averaging is of interest.

Introduction

The primary outcome in cancer randomised controlled trials (RCTs) is often an intermediate outcome such as progression-free survival, while a health technology assessment (HTA) involves estimating mean overall survival (OS). The duration of an RCT is often short relative to the lifetime of participants and relatively few deaths may occur culminating in a type of data gap referred to as immature data.[1] An estimate of mean OS involves specifying a statistical model for the data-generating process and estimating the parameters in the model. There will be structural uncertainty about the true model with different choices giving rise to different estimates of survival functions and mean OS.

The choice of which model to use for decision-making involves: 1) proposing a single model and deciding whether it is adequate for the purpose intended; 2) proposing a range of possible models and making comparisons of the extent to which inferences change according to each model; or 3) creating a single model involving features of all models that are considered plausible using model averaging.[2] In England, submissions to the National Institute for Health and Care Excellence (NICE) often utilise clinical experts who are asked to express their beliefs about the plausibility of different models and to state which model is clinically most plausible.

The aim of this paper is to contrast the retrospective and prospective use of experts' beliefs in choosing between survival models in economic evaluations. We present a case study to illustrate the prospective use of experts and provide a discussion.

**Methods**

Further details are provided in Appendix S1.

*Analytical Framework*

NICE Decision Support Unit Technical Support Document 14 says, "uncertainty [about parameters in survival models] can be taken into account using the variance-covariance matrices for the different parametric models".[3] This approach to dealing with uncertainty implies estimating parameters approximately either using frequentist methods, the predominant framework used in HTA, or by plugging posterior means, standard deviations and covariances from a Bayesian analysis into a multivariate normal distribution.[4] However, frequentist methods do not generate a posterior joint distribution of parameters as required in an economic model and do not permit the inclusion of external information in the form of experts' beliefs expressed as a probability distribution.

*The Bayesian method*

We assume that we have data from an RCT comparing experimental and control treatments. Furthermore, we assume that we have $M$ alternative models, which typically include members of the Generalised $F$ distribution[5] and the Gompertz distribution but could include more flexible models. Each model, $m = 1, \ldots, M$, comprises a likelihood function, $f_m(\mathcal{D}|\theta_m)$, indexed by parameters $\theta_m \in \Theta_m$ having prior distribution $f_m(\theta_m)$. The prior distribution represents information that is available prior to (or, more generally, in addition to) the data, $\mathcal{D}$. In the case of experts' beliefs there is no temporal requirement, only that the experts are not aware of the data or results of any analysis of them before expressing their prior beliefs to avoid the possibility of using the evidence twice.

*Experts' Beliefs*

    *Retrospective beliefs*

The choice of model used for decision-making often involves presenting clinical experts with $M$ model-based estimates of survival functions for each treatment and asking them to express their beliefs about which model is the most plausible. The experts' judgements might be obtained as follows:

- *Looking at the extrapolated survival functions, do the proportions of participants surviving after t months appear to be clinically plausible?*

- *Do any of the survival functions seem to reflect clinical practice?*

  - *If not, what would a clinically plausible extrapolated survival function look like?*

  - *If so, which survival function do you think is clinically most plausible?*

In principle, there is nothing wrong with asking experts to express their retrospective judgements about the weight to be given to different models, although it is not usually a sensible approach and is not a good way to reach posterior judgements. To do so is asking experts to think about both how well the models fit the data and how plausible they were *a priori*. Posterior judgements will almost invariably be better when the two sources of information – prior information and sample data – are considered independently.

The typical process used in HTA does not require experts to provide a rationale for their judgements or to quantify their retrospective beliefs according to a formal elicitation process. In addition, experts are rarely provided with a measure of uncertainty associated with survival functions. Consequently, experts' retrospective judgements could be consistent with several models for the data-generating process irrespective of what the model-based central estimates imply.

Furthermore, asking experts to express their beliefs about the proportion of participants expected to survive after *t* months in clinical practice is potentially misleading. The quantities in the target population and the population defined by the entry criteria of the RCT are not necessarily the same because they depend on the joint distribution of prognostic factors in the respective populations.

*Prospective beliefs*

In a study in which a relatively large proportion of participants experience the event, the data will provide the majority of the evidence about parameters and will dominate any reasonable prior beliefs. Many Bayesian analyses make use of independent, so-called non-informative prior distributions for model parameters representing a state of prior ignorance.[6] When data is immature posterior joint distributions for model parameters and functions of them will not represent reasonable posterior beliefs unless prior joint distributions for them represent reasonable prior beliefs.[7]

*Interpreting clinical experts' retrospective choice of model*

We begin by assuming that we perform a Bayesian analysis in which the prior joint distribution for model parameters represents the experts' beliefs if we had been able to elicit it without reference to the data. After being presented with model-based (posterior) survival functions experts are asked to assess them relative to their (posterior) beliefs. Thus, the current process effectively produces $M + 1$ posterior survival functions for each treatment each purporting to be the experts' posterior survival functions. However, experts are not perfect at being able to assimilate information and it is reasonable for someone to make different judgements about an uncertain quantity when they think about the evidence in different ways. Nevertheless, particular estimates will be preferred through being from a more formal analysis. The preferred estimates are Bayesian posterior estimates because using Bayes' theorem is usually preferred when making direct assessments of posterior probability.

Typically, we do not have Bayesian posterior estimates but have frequentist estimates. If we assume that prior information about model parameters *before observing the data* would have been weak then the $M$ frequentist estimates of survival functions, $S_m(t)$, *approximate* to the experts' Bayesian posterior survival functions. Hence, the analysis provides $M$ posterior estimates of the proportion of participants alive over the observed and extrapolated periods. If the posterior estimates are not close to the experts (usually unquantified posterior) beliefs then, on the basis that experts are not perfect at assimilating information, they might wish to alter their (posterior) beliefs so that the mean of their distribution match the estimates. However, there are $M$ estimates of survival functions and it is not obvious how the experts discriminate between models while accounting for parameter and structural uncertainty.

*Elicitation of expert's beliefs about model parameters*

It is difficult for experts to express beliefs about parameters in survival models; it is better to ask their opinions about observable quantities.[8-13] Furthermore, it is important to acknowledge that beliefs about parameters within and across survival models are correlated in the sense that beliefs about the true value of one parameter affect beliefs about the true value of another parameter. This is done through elaboration by specifying a one-to-one transformation of the model parameters such that the quantities elicited are assumed to be independent.[14] Observable quantities could be the proportion of participants out of the whole population defined by a study's inclusion/exclusion criteria who survive, $S(t_0)$, at some time, $t_0$, and functions of that quantity at one or more times depending on the number of parameters to be estimated. Beliefs about functions of $S(t_0)$ could be expressed as risk differences or relative risks. Parameter constraints are required to ensure that the process does not generate proportions less than zero or greater than one, to ensure that survival functions are monotonically decreasing and that it generates clinically plausible summaries.

*Model comparison*

   *Information criterion*

Several relative goodness-of-fit measures are available to compare different models and each incorporate a different bias correction.[15, 16] Of these, Akaike's Information Criterion (AIC) and Bayesian Information Criterion (BIC) are the ones most commonly used in HTA with smaller values indicating a better fitting model.[17] AIC identifies the model with the best predictive ability based on Kullback-Leibler divergence:

$$\text{AIC}_m = -2 \times \log[f_m(\mathcal{D}|\hat{\theta}_m)] + 2 \times p_m$$

where $\hat{\theta}_m$ is the maximum likelihood estimate of $\theta_m$ and $p_m$ represents the number of parameters in model $m$. BIC is a function of the data only and was derived as a large-sample approximation to twice the logarithm of the Bayes factor when comparing two competing models:[18]

$$\text{BIC}_m = -2 \times \log[f_m(\mathcal{D}|\hat{\theta}_m)] + p_m \log(n_m),$$

where $n_m$ is the number of observations. Volinsky et al.[19] defined the prior distribution that is implicit when using BIC as the unit-information prior, which is vague, based on information from one observation, and proper. Furthermore, they recommend that the penalty for model complexity in the case of time-to-event data with censored observations be the number of uncensored observations. AIC and BIC are based on different underlying principles and can produce different conclusions with BIC based on a bigger adjustment for model complexity that increases with sample size (or events), $n_m \geq 8$. Model choice based on BIC is consistent as the sample size increases in the sense that if one of $M$ alternative models is the true model then the probability that it is selected converges to one.[20] Information criterion assess the extent to which models with parameters estimated using the data alone provide a good representation of the data but provide no information with respect to extrapolated survival functions[21, 22].

*Bayes factors*

The marginal likelihood for each model, also known as Bayesian model evidence (BME), is $f_m(\mathcal{D}) = \int f_m(\mathcal{D}|\theta_m) f_m(\theta_m) d\theta_m.$[16]

The ratio of marginal likelihoods of two models is the Bayes factor and summarises the relative support for model $i$ versus model $j$:[2, 23]

$$B_{ij} = \frac{f_i(\mathcal{D})}{f_j(\mathcal{D})}$$

Bayes factors depend on prior information about model parameters and explicitly account for experts' beliefs about the proportion of participants expected to survive over the lifetime of participants through the prior joint distribution about model parameters.

*Model probabilities*

Experts cannot state with certainty "*which survival function do you think is clinically most plausible?*". They are effectively asked to consider their posterior beliefs about each model given the observed data, which depend on the prior joint distribution of model parameters through the marginal likelihood for each model:[24]

$$f(m|\mathcal{D}) = \frac{f_m(\mathcal{D})f(m)}{\sum_{m=1}^{M} f_m(\mathcal{D}) f(m)}$$

*Computing model probabilities*

There is no unique way to specify prior ignorance about model probabilities; options depend on the number of parameters in each model and whether models are nested.

Formulating and interpreting weights of similar models, such as members of the Generalised F distribution, is not straightforward because of a phenomenon known as dilution. Dilution arises when models are highly correlated or give similar predictions, and applies to prior and posterior model probabilities.[25-27] Garthwaite et al. defined the following properties:[27]

- The dilution property: Models that are highly correlated should be given smaller weights than those that have a low correlation.

- The strong dilution property: When a new model is added to a set of models and it is identical to at least one other model then the weights should be shared amongst those models, while the weights given to the other models should be unchanged.

- The monotonicity property: When a new model is added, weights of models already in the set will generally change but none of them should increase.

In principle, prior model probabilities could be elicited directly from experts. However, prior joint distributions about model parameters already provide information about the plausibility and similarity of different models. George proposed three dilution prior distributions, including one based on distance measures, $D(m, m')$, between models $m$ and $m'$ that is applicable to any class of model. Hellinger's distance between marginal likelihoods is:[26]

$$D^H(m, m') = \int \left[ f_m^{1/2}(y) - f_{m'}^{1/2}(y) \right]^2 dy$$

where the integration is over the sample space of the data. Prior model probabilities could then be defined as:

$$f_{D^H}(m) \propto a_m$$

where $a_m = \sum_{m'} D^H(m, m')$, the sum of the distances from model $m$ to every other model. More weight will be assigned to distant, isolated models and less weight will be assigned to similar models.

Posterior model probabilities are conditional on the $M$ models, although it is unlikely that any of the proposed models is the true model for the data-generating process. Consequently, the model probabilities are interpreted as weights reflecting the relative importance of each of the $M$ models.[2] Given the phenomenon of dilution, model choice based on the largest posterior model probability should be avoided.

## Case study

We illustrate the methodology using data on recurrence-free survival (RFS) from randomised and non-randomised participants with node-positive breast cancer from the German Breast Cancer Study Group (Appendix S2).[28]

### Results

Suppose that, before the data was available, we elicited the prior quantiles for the uncertain quantities at 5 and 10 years, and fitted the distributions as given in Table 1.

Table 1: Elicited prior quantiles and fitted distributions of uncertain quantities for RFS

|  | $S_1(5)$ | $\delta_{21}(5)$ | $\delta_{11}(10)$ | $\delta_{22}(10)$ |
|---|---|---|---|---|
| Lower quartile | 0.37 | 0.01 | 0.26 | 0.25 |
| Median | 0.40 | 0.05 | 0.30 | 0.30 |
| Upper quartile | 0.45 | 0.10 | 0.35 | 0.37 |
| Distribution | Beta(27.09, 39.58)) | $N(0.053, 0.067^2)$ | Beta(14.51, 33.09) | Beta(8.33, 18.61) |

The prior distribution for the proportion of participants given Treatment 1 surviving to 5 years is equivalent to 66.67 participants of whom 27.09 survived giving a mean proportion surviving of 40.6%.

Visually there is little to distinguish between the prior survival function for a Weibull distribution and other members of the generalised F; each of them coincides exactly with elicited proportions of participants surviving. The Gompertz survival function is most different as a consequence of its hazard function being strictly increasing (Figure 1 and Figures S2.1-4, respectively).

Figure 1: Weibull distribution: Join distributions of model parameters, survival functions and Bayesian model evidence

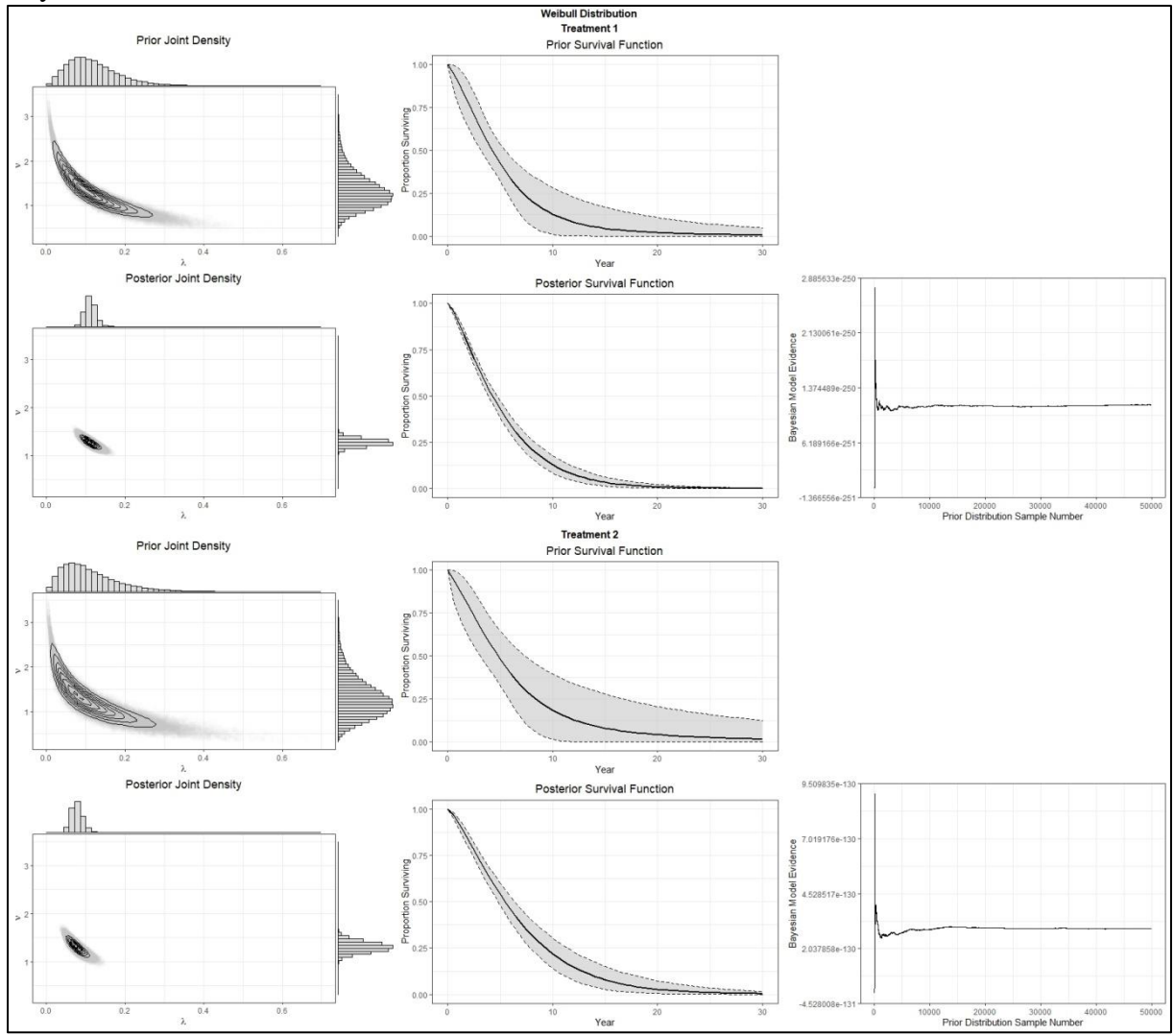

The Hellinger's distances show that (Table 2): for Treatment 1, the Gompertz distribution is 'distant' to all other models, the exponential distribution is also 'distant' to the lognormal and log-logistic distributions and 'close' to the Weibull distribution, and the lognormal distribution is 'close' to the log-logistic distribution; for Treatment 2, the Gompertz distribution is 'distant' to all other models, the exponential distribution is also 'distant' to the lognormal and log-logistic distributions and 'close' to the Weibull distribution, the Weibull distribution is also 'close' to the lognormal distribution, and the lognormal distribution is 'close' to the log-logistic distribution. The prior model weights are not equal with more weight *a priori* given to the Gompertz distribution and less weight given to more similar models.

The empirical hazard functions suggest an increase in the hazard of an event over the initial 18-24 months followed by a decrease in hazard up to approximately 3.5 years. There is a suggestion that the hazard may increase after that, although there is considerable uncertainty and no sample evidence beyond 7 years (Figure 2).

Figure 2: Empirical hazard functions with 95% confidence intervals

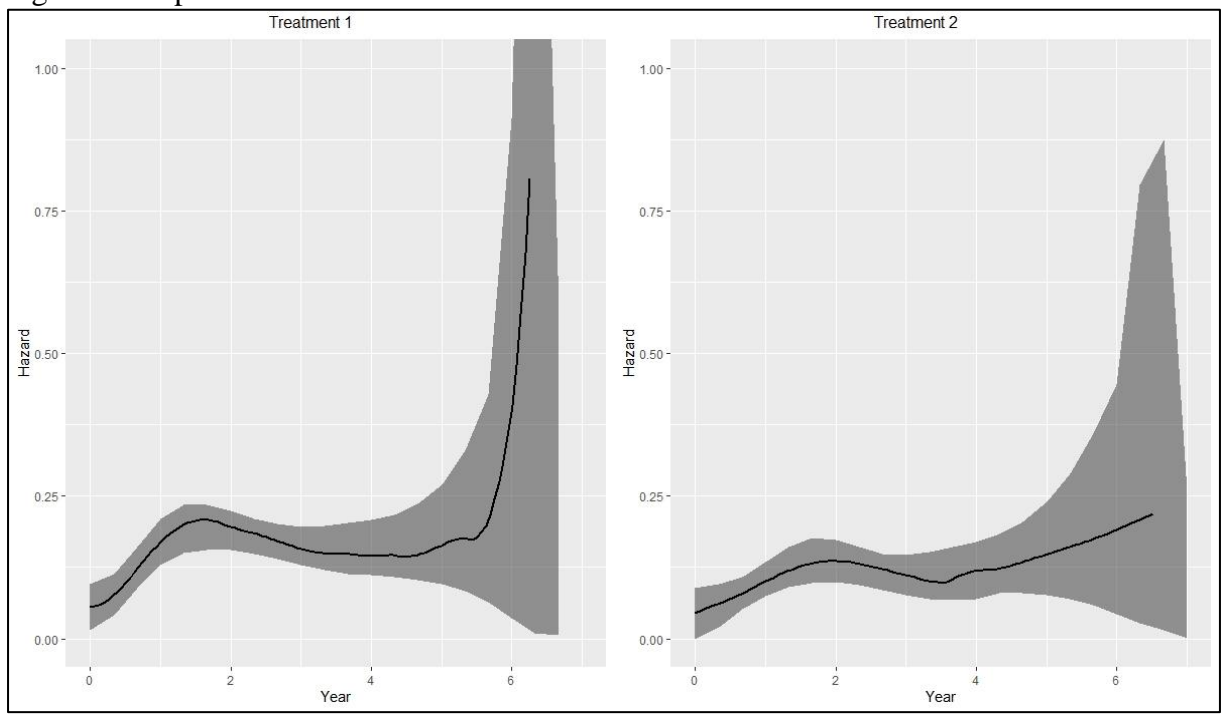

Table 2: Prior and Posterior Model Probabilities

| Treatment 1 | Hellinger's Distance | | | | | $\sum_{m'} D^H(m, m')$ | $f(m)$ | $f_m(\mathcal{D})$ | $f(m\|\mathcal{D})$ |
|---|---|---|---|---|---|---|---|---|---|
| | Exponential | Weibull | Log-logistic | Lognormal | Gompertz | | | | |
| Exponential | 0 | 0.0000095 | 0.0000536 | 0.0000391 | 0.0000453 | 0.0001475 | 0.2288124 | 1.58E-253 | 0.0000000 |
| Weibull | 0.0000095 | 0 | 0.0000207 | 0.0000124 | 0.0000301 | 0.0000727 | 0.1127737 | 1.14E-250 | 0.0000124 |
| Lognormal | 0.0000391 | 0.0000124 | 0.0000027 | 0 | 0.0000519 | 0.0001061 | 0.1645575 | 6.29E-246 | 0.9951537 |
| Log-logistic | 0.0000536 | 0.0000207 | 0 | 0.0000027 | 0.0000570 | 0.0001340 | 0.2078213 | 2.42E-248 | 0.0048339 |
| Gompertz | 0.0000453 | 0.0000301 | 0.0000570 | 0.0000519 | 0 | 0.0001844 | 0.2860351 | 2.46E-254 | 0.0000000 |
| **Treatment 2** | **Hellinger's Distance** | | | | | $\sum_{m'} D^H(m, m')$ | $f(m)$ | $f_m(\mathcal{D})$ | $f(m\|\mathcal{D})$ |
| | Exponential | Weibull | Log-logistic | Lognormal | Gompertz | | | | |
| Exponential | 0 | 0.0000072 | 0.0000374 | 0.0000259 | 0.0000453 | 0.0001158 | 0.2170963 | 1.52E-131 | 0.0006924 |
| Weibull | 0.0000072 | 0 | 0.0000143 | 0.0000082 | 0.0000288 | 0.0000584 | 0.1095711 | 2.94E-130 | 0.0067560 |
| Lognormal | 0.0000259 | 0.0000082 | 0.0000018 | 0 | 0.0000469 | 0.0000828 | 0.1553155 | 2.82E-128 | 0.9204097 |
| Log-logistic | 0.0000374 | 0.0000143 | 0 | 0.0000018 | 0.0000509 | 0.0001043 | 0.1956360 | 1.69E-129 | 0.0694037 |
| Gompertz | 0.0000453 | 0.0000288 | 0.0000509 | 0.0000469 | 0 | 0.0001719 | 0.3223811 | 4.05E-131 | 0.0027392 |

Of the models considered, the lognormal and log-logistic distributions are the only ones associated with increasing followed by decreasing hazards, and none of the considered models allow for multiple turning points.

Of the models considered, the Bayes factors indicate decisive evidence that the data from Treatment 1 was generated from a lognormal distribution rather than from any other of the models. The Bayes factors indicate decisive evidence that the data from Treatment 2 was generated from a lognormal distribution in preference to exponential and Gompertz distributions, very strong evidence in preference to a Weibull distribution, and strong evidence in preference to a log-logistic distribution. The posterior model weights could be used to create a single model if formal model averaging is of interest.[2, 22, 29, 30]

The Kaplan-Meier and posterior expected parametric survival functions are presented in Figure 3.

Figure 3: Kaplan-Meier and posterior expected parametric survival functions

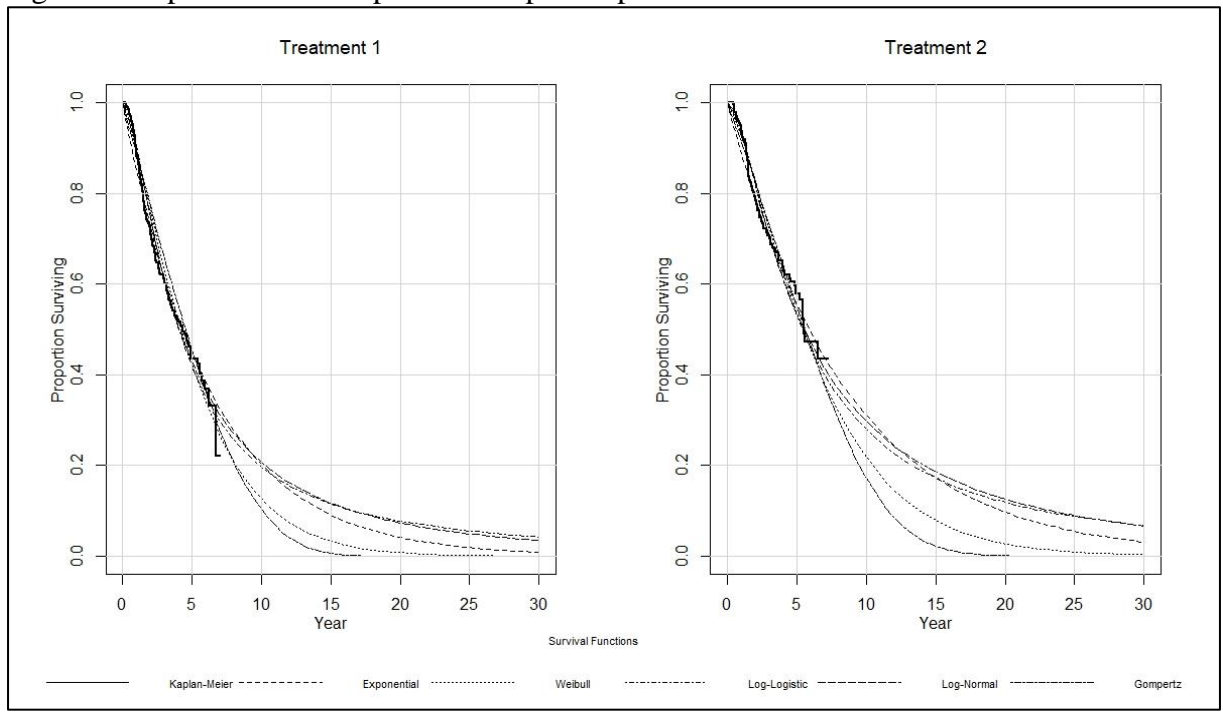

The lognormal distribution suggests a posterior incremental mean RFS of 3.04 years (95% CrI: -0.00, 7.34) (Table 3). The log-logistic distribution suggests a similar posterior incremental mean RFS of 3.73 years but with more uncertainty (95% CrI: -1.42, 12.13).

Table 3: RFS: Summaries of prior and posterior means

| Distribution [a] | Prior Mean (yrs) (95% CrI) | | Posterior Mean (yrs) (95% CrI) | | Posterior Incremental Mean (yrs) 95% CrI |
|---|---|---|---|---|---|
| | Treatment 1 | Treatment 2 | Treatment 1 | Treatment 2 | |
| Exponential | 5.64 (4.08, 7.77) | 6.69 (3.97, 11.06) | 6.27 (5.56, 7.08) | 8.58 (7.17, 10.28) | 2.31 (0.79, 4.05) |
| Weibull | 5.47 (4.16, 8.37) | 6.77 (4.27, 13.60) | 5.26 (4.71, 5.93) | 6.74 (5.69, 8.18) | 1.48 (0.29, 2.97) |
| Lognormal | 6.14 (4.53, 10.87) | 8.00 (4.64, 20.68) | 7.51 (6.24, 9.20) | 10.54 (7.83, 14.80) | 3.04 (-0.00, 7.34) |
| Log-logistic | 7.02 (4.65, 16.72) | 9.19 (4.75, 28.63) | 9.21 (7.17, 12.35) | 12.94 (8.71, 21.23) | 3.73 (-1.42, 12.13) |
| Gompertz | 5.93 (5.01, 7.17) | 6.65 (5.16, 9.07) | 5.10 (4.69, 5.57) | 6.06 (5.35, 6.94) | 0.96 (0.20, 1.85) |

a: Survival functions for distributions with inherently heavy tails have greater proportions of patients surviving up to 50 years, the maximum mean survival used as a constraint in the analysis.

Discussion

Clinical experts are often asked to retrospectively validate [sic] the choice of survival function used for decision making, thereby appearing to provide credibility to its choice. A problem that is often ignored with this approach arises because they are asked to express judgements in terms of what they expect in clinical practice rather than with respect to the population defined by a study. The target population is relevant but only in defining the baseline survival function on which to project estimates of relative treatment effect. In some situations the population defined by one or more studies may be representative of the target population but, if not, it may be necessary to use a different source of evidence to generate the baseline survival function. The impact on reimbursement decisions of using evidence from clinical trials to estimate both the relative treatment effect and the baseline survival function of the target population is unknown, particularly when decisions are based on a single study.

The *elicitation process* itself is often not transparent and flawed. Hindsight is a wonderful thing and experts appear to have little difficulty expressing their retrospective (posterior) beliefs about the proportion of participants surviving to different times after seeing model-based estimates. In practice, better posterior judgements would be obtained by considering prior information and sample data separately and applying Bayes' Theorem. Cope et al. demonstrated the feasibility of incorporating experts' beliefs in a fractional polynomial survival model but with some limitations.[31] Experts were mainly from the study network and were aware of the results of the study which were published in the year prior to the elicitation so were effectively expressing posterior rather than prior beliefs.

Prior information always exists and should be used to strengthen inferences and improve decision-making. In the context of a survival analysis, and in addition to any external evidence available, it involves omitting implausible parameter sets, ensuring that prior survival functions are monotonically decreasing and that prior estimates of population mean are bounded. This information alone can strengthen inferences relative to a conventional analysis. Experts' judgements should be elicited as probability distributions using a justifiable and transparent process before observing the data. Experts should be aware of the same body of evidence when making their judgements and have an opportunity to discuss amongst them the basis for their judgements. The process followed and the elicited distributions must be specified in a protocol to facilitate review by decision-makers, referees and other stake-holders.

We have shown for five standard parametric models how elaboration about observable quantities induces correlation between parameters in survival models. The uncertain quantities of interest were functions of the proportion of participants that survive out of the whole population defined by a study's inclusion/exclusion criteria at one or more times depending on the number of parameters. The choice of times at which to elicit beliefs is specified as part of the elicitation process. Consequently, the prior joint distribution of model parameters is not unique. We assumed the same standard parametric model for each treatment arm, which is common practice in submissions to NICE in the UK. It is not necessary to assume the same model for each treatment arm because the elicited quantities do not depend on the model for the data although, for standard models, not doing so would make it difficult to combine evidence across multiple studies. We made no assumption of a constant treatment effect (e.g. proportional hazard, proportional odds or acceleration factor) which we consider an unnecessary and often implausible modelling assumption.

Specifying plausible survival models is not straightforward and requires clinical knowledge of the underlying disease process, risk factors, the mechanisms of action of different treatments and the risk of an event over time. Such models may include mixture models that model heterogeneous time-to-event data, change-point models that reflect changes in hazard functions over time because of changes in treatment (e.g. induction and maintenance phases) and more flexible models that capture more complex features. Although models that are members of the Generalised $F$ distribution are commonly used in HTAs, they are likely at best to provide only an approximation to the lifetime hazard functions.[5]

Information criterion using sample data alone cannot tell which model is the most appropriate data-generating process during the extrapolation period. External information should be used to strengthen inferences. When interest is in making inferences with respect to a preferred model, this should be based on more formal criteria such as Bayes factors rather than asking experts to choose a model.

We make no assertion regarding the appropriateness or use of model averaging. However, when there is interest in model averaging, it is important to appreciate that so-called non-informative prior information about model probabilities can be misleading because of the phenomenon known as dilution. Furthermore, information criterion provide an often poor approximation to BME. We have shown that exact estimation of BME is feasible and computationally inexpensive.

## Conclusion

Retrospective use of experts' beliefs to validate (sic) a model is potentially misleading, may not achieve its intended objective and is an inefficient use of information. Experts' posterior beliefs should be generated using Bayes' theorem by combining prospectively elicited prior beliefs with sample data. This can be used to strengthen inferences, facilitate the choice of model and mitigate the impact of dilution on model probabilities in situations when model averaging is of interest. Further research is required to develop guidance on the elicitation of experts' beliefs about parameters in standard and other survival models, the generation of practical examples and formulating prior model probabilities.

References

1. Tai T-A, Latimer N, Benedict A, Kiss Z, Nikolaou A. Prevalence of immature survival data for anti-cancer drugs presented to the National Institute for Health and Care Excellence and impact on decision making. *Value in Health* 2021;24:505-12.
2. O'Hagan A, Forster J. Kendall's Advanced Theory of Statistics. Bayesian Inference. Second Edition.; 2004.
3. Latimer N. NICE DSU Technical Support Document 14: Undertaking survival analysis for economic evaluations alongside clinical trials - extrapolation with patient-level data. 2011.
4. Gallacher D, Auguste P, Connock M. How do pharmaceutical companies model survival of cancer patients? A review of NICE Single Technology Appraisals in 2017. *International Journal of Technology Assessment in Health Care* 2019;35:160-7.
5. Cox C. The generalised F distribution: An umbrella for parametric survival analysis. *Statistics in Medicine* 2008;27:4301-12.
6. Kass R, Wasserman L. The selection of prior distributions by formal rules. *JAMA* 1996;91:1343-70.
7. Stevens J. Using evidence from randomised controlled trials in economic models: what information is relevant and is there a minimum amount of sample data required to make decisions? *Pharmacoeconomics* 2018;36:1135-41.
8. Kadane J, Wolfson L. Experiences in elicitation. *The Statistician* 1998;47:3-19.
9. Singpurwalla N. An interactive PC-based procedure for reliability assessment incorporating expert opinion and survival data. *Journal of the American Statistical Association* 1988;83:43-51.
10. Berger J, Sun D. Bayesian analysis for the poly-Weibull distribution. *Journal of the American Statistical Association* 1993;88:1412-8.
11. Kaminskiy M, Krivtsov V. A simple procedure for Bayesian estimation of the Weibull distribution. *IEEE Transactions on Reliability* 2005;54:612-6.
12. Bousquet N. A Bayesian analysis of industrial lifetime data with Weibull distributions. [Research Report] RR-6025, INRIA. pp. 24 inria-00115528v4; 2006.
13. Ren S, Oakley J. Assurance calculations for planning clinical trials with time-to-event outcomes. *Statistics in Medicine* 2014;33:31-45.
14. O'Hagan A. Expert knowledge elicitation: subjective but scientific. *The American Statistician* 2019;73:69-81.
15. Emiliano P, Vivanco M, de Menezes F. Information criteria: How do they behave in different models? . *Computational Statistics and Data Analysis* 2014;69:141-53.
16. Schoniger A, Wohling T, Samaniego L, Nowak W. Model selection on solid ground: Rigorous comparison of nine ways to evaluate Bayesian model evidence. *Water Resources Research* 2014;50:9484-953.
17. Bell-Gorrod H, Kearns B, Stevens J, Thokala P, Labeit A, Latimer N*, et al.* A review of survival analysis methods used in NICE technology appraisals of cancer treatments: consistency, limitations and areas for improvement. *Medical Decision Making* 2019;39:899-909.
18. Schwarz G. Estimating the dimension of a model. *Annals of Statistics* 1978;6:461-4.
19. Volinsky C, Raftery A. Bayesian information criterion for censored survival models. *Biometrics* 2000;56:256-62.
20. Jackson C, Thompson S, Sharples L. Accounting for uncertainty in health economic decision models by using model averaging. *Journal of the Royal Statistical Society Series A* 2009;72:383-404.
21. Bagust A, Beale S. Survival analysis and extrpolation modelling of time-to-event clinical trial data for economic evaluation: An alternative approach. *Medical Decision Making* 2014;34:343-51.

22. Negrin M, Nam J, Briggs A. Bayesian solutions for handling uncertainty in survival extrapolation. *Medical Decision Making* 2017;37:367-76.
23. Kass R, Raftery A. Bayes Factors. *Journal of the American Statistical Association* 1995;90:773-95.
24. Clyde M. Model uncertainty and health effect studies for particulate matter. *Environmetrics* 2000;11:745-63.
25. George E. Invited discussion of "Bayesian model averaging and model search strategies" by M.A. Clyde: Oxford: Oxford University Press.; 1999.
26. George E. Dilution prior: Compensating for model space redundancy. . *IMS Collections* 2010;6:158-65.
27. Garthwaite P, Mubwandarikwa E. Selection of weights for weighted model averaging. *Australian & New Zealand Journal of Statistics* 2010;52:363-82.
28. Sauerbrei W, Royston P, Bojar H, Schmoor C, Schumacher M. Modelling the effects of standard prognostic factors in node-positive breast cancer. *British Journal of Cancer* 1999;79:1752-60.
29. Hoeting J, Madigan D, Raftery A, Volinsky C. Bayesian model averaging: A tutorial. *Statistical Science* 1999;4:382-47.
30. Thamrin S, McGree J, Mengersen K. Modelling survival data to account for model uncertainty: a single model or model averaging? *SpringerPlus* 2013;2:665.
31. Cope A, Ayers D, Zhang J, Batt K, Jansen J. Integrating expert opuinion with clinical trial data to extraopolate long-term survival: a case study of CAR-T therapy for children and young adults with relapsed or refractory acute lymphoblastic leukemia. *BMC Medical Research Methodology* 2019;19.

# Appendix S1: The Bayesian method

## Estimating parameters

To allow for uncertainty about the relative plausibility of each model, we express the likelihood as $f(\mathcal{D}|m, \theta_m)$. A Bayesian analysis requires a prior joint distribution for $(m, \theta_m)$, which can be factorised as:

$$f(m, \theta_m) = f(\theta_m|m)f(m).$$

After observing data, $\mathcal{D}$, the posterior joint distribution, $f(m, \theta_m|\mathcal{D})$, is:

$$f(m, \theta_m|\mathcal{D}) = f(\theta_m|m, \mathcal{D})f(m|\mathcal{D})$$

$$= \frac{f(\theta_m|m, \mathcal{D})f_m(\mathcal{D})f(m)}{\sum_{m=1}^{M} f_m(\mathcal{D})f(m)}.$$

## Marginal likelihood

Schöniger et al evaluated the performance of different information criterion as an approximation to $f_m(\mathcal{D})$ and concluded that they "are often heavily biased and that the choice of approximation method substantially influences the accuracy of model ranking. For reliable model selection, bias-free numerical methods should be preferred over [information criterion] whenever computationally feasible".[1]

## Elicitation of experts' beliefs

### *Process*

The NICE Guide to the Methods of Technology Appraisal 2013 acknowledges that probability distributions "should not be arbitrarily chosen" and that "formal elicitation methods are available if there is a lack of [sample] data" with which to generate joint distributions of parameters.[2] There is an extensive literature on the elicitation of experts' beliefs using different elicitation protocols namely behavioural aggregation using the Sheffield Elicitation Framework (SHELF) (i.e. face-to-face elicitation), mathematical aggregation using Cooke's method (i.e. individual elicitation), and the Delphi method using a mixture of behavioural and mathematical aggregation.[3-7] There are advantages and disadvantages associated with each approach although, when there are multiple experts, we prefer SHELF, which generates a prior distribution that represents the belief of a rationale impartial observer and not necessarily a consensus distribution. A feature of SHELF is the generation of an evidence dossier containing all relevant information relating to the uncertain quantities of interest. The evidence dossier ensures that judgements are based on the same body of evidence and mitigates any bias associated with the availability heuristic.

Probability distributions can be elicited using fixed or variable interval methods such as the trial roulette method[8,9] and bisection method[9], respectively, and fitted using the freely available SHELF software[10] or the online tool MATCH.[9]

Eliciting beliefs about an odds ratio would avoid generating implausible proportions, although an odds ratio is unlikely to be a quantity that an expert is able to express beliefs about in most practical situations.

*Probability distributions*

Table S1.1: Probability distributions

| Distribution | Density Function<br>$f(t\mid \cdot)$ | Survival Function<br>$S(t\mid \cdot)$ | Hazard Function<br>$h(t\mid \cdot) = f(t\mid \cdot)/S(t\mid \cdot)$ |
|---|---|---|---|
| Exponential | $f(t\mid\lambda) = \lambda e^{-\lambda t}$<br>$t > 0;$ rate (scale) $\lambda > 0$ | $S(t\mid\lambda) = e^{-\lambda t}$ | $h(t\mid\lambda) = \lambda$<br>Constant |
| Weibull | $f(t\mid\nu,\lambda) = \nu\lambda t^{\nu-1} e^{-\lambda t^{\nu}}$<br>$t > 0;$ shape $\nu > 0;$ scale $\lambda > 0$ | $S(t\mid\nu,\lambda) = e^{-\lambda t^{\nu}}$ | $h(t\mid\nu,\lambda) = \nu\lambda t^{\nu-1}$<br>$\nu = 1$: Constant – exponential distribution<br>$\nu < 1$: Decreasing monotonically<br>$\nu > 1$: Increasing monotonically |
| Lognormal | $f(t\mid\mu,\sigma) = \dfrac{1}{t\sigma\sqrt{2\pi}}\exp\left(-\dfrac{1}{2\sigma^2}(\log(t) - \mu)^2\right)$<br>$t > 0;$ location $\mu \in (-\infty, \infty);$ shape $\sigma > 0$ | $S(t\mid\mu,\sigma) = \phi\left(-\dfrac{\log(t) - \mu}{\sigma}\right)$ | $h(t\mid \cdot) = f(t\mid \cdot)/S(t\mid \cdot)$<br>Unimodal - Increasing then decreasing |
| Log-logistic[a] | $f(t\mid\beta,\theta) = \dfrac{\dfrac{\beta}{\theta}\left(\dfrac{t}{\theta}\right)^{\beta-1}}{\left\{1 + \left(\dfrac{t}{\theta}\right)^{\beta}\right\}^2}$<br>$t > 0;$ shape $\beta > 0;$ scale $\theta > 0$ | $S(t\mid\beta,\theta) = \dfrac{1}{1 + \left(\dfrac{t}{\theta}\right)^{\beta}}$ | $h(t\mid\beta,\theta) = \dfrac{\dfrac{\beta}{\theta}\left(\dfrac{t}{\theta}\right)^{\beta-1}}{1 + \left(\dfrac{t}{\theta}\right)^{\beta}}$<br>$\beta > 1$: Unimodal - Increasing then decreasing<br>$0 < \beta \leq 1$: Decreasing monotonically |
| Gompertz | $f(t\mid\lambda,\theta) = \lambda e^{\theta t}\exp\left(-\dfrac{\lambda}{\theta}(e^{\theta t} - 1)\right)$<br>$t > 0;$ shape $\theta \geq 0;$ scale $\lambda > 0$ | $S(t\mid\lambda,\theta) = \exp\left(-\dfrac{\lambda}{\theta}(e^{\theta t} - 1)\right)$ | $h(t\mid\lambda,\theta) = \lambda e^{\theta t}$<br>$\theta > 0$: Increasing monotonically from $\lambda$ at time $t = 0$<br>$\theta = 0$: Constant – exponential distribution |

[a] The mean is not defined when $\beta \leq 1$

The exponential, gamma, Weibull, lognormal and log-logistic distributions are nested models and are all members of the Generalised F distribution (Figure S1.1).[11]

Figure S1.1: Generalised F distribution

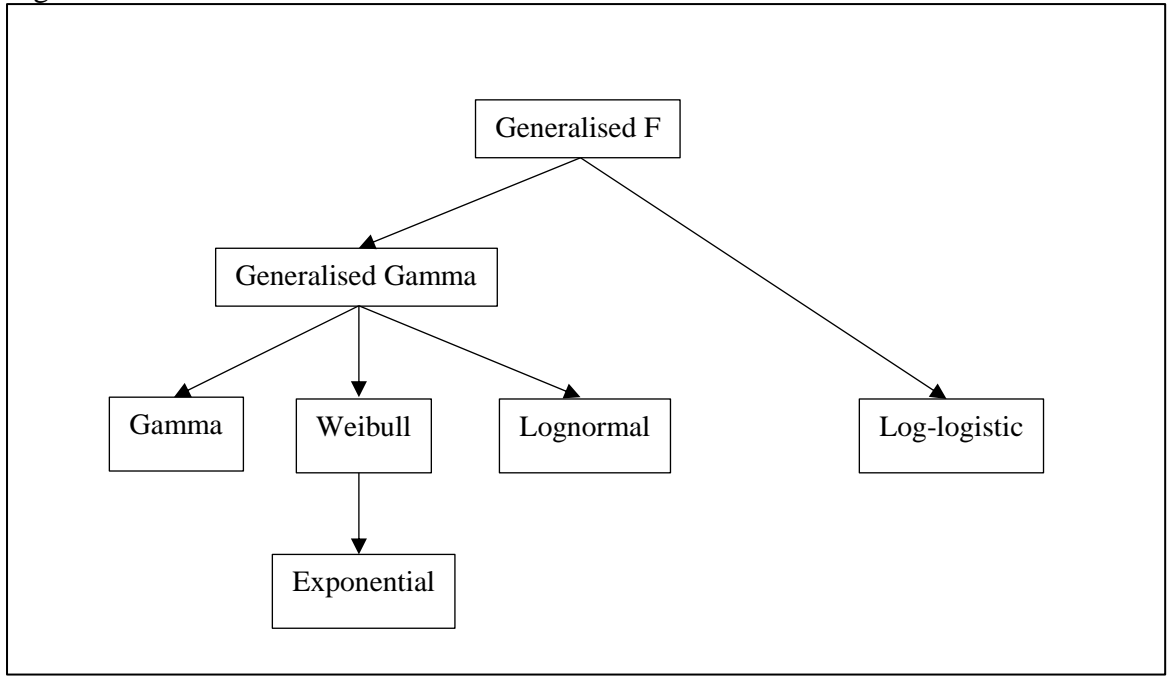

A Gompertz distribution is appropriate when the hazard increases over time so that $S(t) \to 0$ as $t \to \infty$; a negative value of $\theta$ implies that a proportion of people are immortal; a negative value of $\lambda$ implies that the hazard of an event is negative until $\lambda > e^{\theta}$. In practice, values of $\lambda$ and $\theta$ are generally restricted to a limited sample space and are highly correlated; the smaller the value of $\theta$, the larger is the value of $\lambda$. The mode of a Gompertz distribution is:[12]

$$\text{Mode}[X] = \frac{1}{\theta}\log\left(\frac{\theta}{\lambda}\right)$$

Hence, the mode is negative when $\theta < \lambda$ and is zero when $\theta = \lambda$, which implies that $\theta$ must be greater than $\lambda$ for a plausible survival model.

The log-logistic and Gompertz distributions are not available as standard distributions in JAGS. Time-to-event data can be modelled using these distributions in JAGS by making use of the "zeros trick". A Poisson($\phi$) observation of zero has likelihood $\exp(\phi)$ so that if observed data is a set of 0's and $\phi[t_i]$ is set to $-\log(L[t_i])$, we obtain the correct likelihood contribution.[13]

For time-to-event data, the likelihood is:

$$L(t_i) = \prod_{i=1}^{n} f(t_i|\theta)^{d_i} S(t_i|\theta)^{(1-d_i)}.$$

$$\Rightarrow \log[L(t_i)] = \sum_{i=1}^{n} \left(d_i \times \log(f(t_i|\theta)) + (1-d_i) \times \log(S(t_i|\theta))\right)$$

$$= \sum_{i=1}^{n} d_i \times \log(h(t_i|\theta)) + \log(S(t_i|\theta))$$

where $d_i$ is an indicator variable that equals 1 if the event is observed and 0 if the event is censored.

*Exponential distribution: Exact prior distribution for the rate parameter for a single treatment*

We derive the prior distribution for the rate parameter $\lambda$ of an exponential distribution based on experts' beliefs about the proportion of patients out of the whole population defined by a study's inclusion/exclusion criteria who survive beyond time $t$ expressed using a beta distribution.

Let $\pi = S(t) = P(T > t) = \exp(-\lambda t)$ represent the probability of surviving beyond time $t$.

Re-arranging with respect to $\lambda$ gives $\lambda = -\frac{\log(\pi)}{t}$.

Taking derivatives, $d\lambda = -\frac{1}{t\pi} d\pi$, so that $\frac{1}{d\lambda} = \frac{-t\exp(-\lambda t)}{d\pi}$.

Let $\pi \sim Beta(\alpha, \beta)$.

Now,

$$F_\Lambda(\lambda) = P(\Lambda < \lambda)$$

$$= P\left(-\frac{\log(\Pi)}{t} < \lambda\right)$$

$$= P(\Pi > \exp(-t\lambda))$$

$$= \int_{\exp(-t\lambda)}^{1} \frac{\Gamma(\alpha+\beta)}{\Gamma(\alpha)\Gamma(\beta)} \pi^{\alpha-1}(1-\pi)^{\beta-1} d\pi.$$

Therefore,

$$f_\Lambda(\lambda) = \frac{d}{d\lambda} F_\Lambda(\lambda)$$

$$= -\frac{t\exp(-\lambda t)}{d\pi} \int_{\exp(-t\lambda)}^{1} \frac{\Gamma(\alpha+\beta)}{\Gamma(\alpha)\Gamma(\beta)} \pi^{\alpha-1}(1-\pi)^{\beta-1} d\pi$$

$$= t\frac{\Gamma(\alpha+\beta)}{\Gamma(\alpha)\Gamma(\beta)}(exp(-\lambda t))^{\alpha-1}(1-exp(-\lambda t))^{\beta-1}\exp(-\lambda t).$$

Knowing the exact form of the prior distribution is not necessary nor possible when interest is in expressing the prior joint distribution of multiple rate parameters from multiple treatments.

*Exponential distribution: Joint prior distributions for the rate parameters of multiple treatments*

The exponential distribution is the simplest parametric model used to analyse time-to-event data. It has a single parameter, $\lambda > 0$, representing the (hazard) rate of an event.

For a control and experimental treatment, the task is to generate a prior joint distribution for the rate parameters, $\lambda_1$ and $\lambda_2$.

Let $S_i(t_0)$ represent the proportion of patients out of the whole population defined by a study's inclusion/exclusion criteria who survive at time $t_0$ on Treatment $i$, $i = 1,2$.

Applying the complimentary log-log transformation to the survival function:

$$\log\left(-\log(S_i(t_0))\right) = \log(\lambda_i) + \log(t_0)$$

The uncertain parameters of an exponential distribution are expressed as functions of $S_i(t_0)$:

$$\lambda_i = \frac{-\log(S_i(t_0))}{t_0}$$

Therefore, rather than eliciting beliefs about the joint distribution, $\pi(\lambda_1, \lambda_2)$, we elicit beliefs about $S_i(t_0)$.

To induce correlation between $S_i(t_0)$, we can elicit beliefs about the following functional parameters and assume them to be independent:[14]

$$S_1(t_0)$$

$$\delta_{21} = S_2(t_0) - S_1(t_0)$$

Beliefs about $S_1(t_0)$ can be represented using beta distributions. Beliefs about $\delta_{21}$ can be represented using a normal distribution or as a scaled beta distribution to allow for the possibility of asymmetry in the prior distribution.

The necessary constraints to ensure plausible parameter values are:

$$S_1(t_0) + \delta_{21} > 0$$

$$1 - S_1(t_0) - \delta_{21} > 0$$

The expected value of an exponential distribution is:

$$E[X] = \frac{1}{\lambda}$$

The median of an exponential distribution is:

$$Med[X] = \frac{\log(2)}{\lambda}$$

*Weibull distribution: Joint prior distributions for the shape and scale parameters of multiple treatments*

There are at least two parameterisations of the Weibull distribution; we use the parameterisation used by OpenBUGS/WinBUGS.[13] and JAGS[15]

For a control and experimental treatment, the task is to generate a prior joint distribution for the shape and scale parameters, $v_1$, $\lambda_1$, $v_2$, and $\lambda_2$. Let $S_i(t_0)$ and $S_i(t_1)$ represent the proportion of participants out of the whole population defined by a study's inclusion/exclusion criteria who survive at times $t_0$ and $t_1$ on Treatment $i$, $i = 1,2$, where $t_1 > t_0$. Applying the complementary log-log transformation to the survival function:

$$\ln\left(-\ln\left(S_i(t_0)\right)\right) = \ln(\lambda_i) + v_i \ln(t_0)$$

$$\ln\left(-\ln\left(S_i(t_1)\right)\right) = \ln(\lambda_i) + v_i \ln(t_1)$$

The parameters of the Weibull distributions are expressed as functions of $S_i(t_0)$ and $S_i(t_1)$:

$$v_i = \frac{\ln\left(\frac{\ln(S_i(t_1))}{\ln(S_i(t_0))}\right)}{\ln\left(\frac{t_1}{t_0}\right)}$$

$$\lambda_i = \frac{-\ln(S_i(t_0))}{t_0^{v_i}}$$

Therefore, rather than eliciting beliefs about the joint distribution, $\pi(v_1, \lambda_1, v_2, \lambda_2)$, we elicit beliefs about $S_i(t_0)$ and $S_i(t_1)$. To induce correlation between $S_i(t_1)$ and $S_i(t_0)$, we can elicit beliefs about the following functional parameters and assume them to be independent:[14]

$$S_1(t_0)$$

$$\delta_{11} = S_1(t_0) - S_1(t_1)$$

$$\delta_{21} = S_2(t_0) - S_1(t_0)$$

$$\delta_{22} = S_2(t_0) - S_2(t_1)$$

Beliefs about $S_1(t_0)$, $\delta_{11}$ and $\delta_{22}$ can be represented using beta distributions. Beliefs about $\delta_{21}$ can be represented using a normal distribution or as a scaled beta distribution to allow for the possibility of asymmetry in the prior distribution.

The necessary constraints to ensure plausible parameter values are:

$S_1(t_0) + \delta_{21} > 0$

$1 - S_1(t_0) - \delta_{21} > 0$

$S_1(t_0) - \delta_{11} > 0$

$S_1(t_0) + \delta_{21} - \delta_{22} > 0$

$S_1(t_1) > 0$

$S_2(t_1) > 0$

Constraints could be placed on $\lambda_i$ by recognising that $S_i(1) = e^{-\lambda_i}$.

In situations when it is believed that the underlying hazard function is strictly increasing (decreasing) over time then the shape parameter, $\nu_i$, could be constrained to be strictly greater than (less than one), although this would be a strong belief.

Constraints on model parameters do not affect the distributions of the elicited quantities.

The expected value of a Weibull distribution is:

$$E[X] = \lambda^{-1/\nu} \Gamma\left(1 + \frac{1}{\nu}\right)$$

The median of a Weibull distribution is:

$$\text{Med}[X] = \lambda^{-1/\nu} (\log(2))^{1/\nu}$$

*Lognormal distribution: Joint prior distribution for the location and shape parameters of multiple treatments*

A lognormal distribution is characterised by a location parameter $\mu \in (-\infty, \infty)$ and shape parameter $\sigma > 0$ representing the mean and standard deviation on the log scale.

For a control and experimental treatment, the task is to generate a prior joint distribution for $\mu_i$ and $\sigma_i$. Let $S_i(t_0)$ and $S_i(t_1)$ represent the proportion of patients out of the whole population defined by a study's inclusion/exclusion criteria who survive at times $t_0$ and $t_1$ on Treatment $i$, $i = 1,2$, where $t_1 > t_0$.

A lognormal distribution can be written as:

$$\log(t) = \mu + \varepsilon$$

where $\varepsilon \sim N(0, \sigma^2)$.

The cumulative distribution and survival functions are:

$$F[t] = \phi\left(\frac{\log(t) - \mu}{\sigma}\right)$$

$$S[t] = 1 - F[\log(t)]$$

$$= \phi\left(-\frac{\log(t) - \mu}{\sigma}\right)$$

Therefore,

$$-\phi^{-1}[S(t)] = \frac{\log(t) - \mu}{\sigma}$$

Let

$$\gamma_0 = -\frac{\mu}{\sigma}$$

and

$$\gamma_1 = \frac{1}{\sigma}$$

so that

$$\mu = -\frac{\gamma_0}{\gamma_1}$$

and

$$\sigma = \frac{1}{\gamma_1}$$

Then,

$$-\phi^{-1}[S(t)] = \gamma_0 + \gamma_1 \log(t)$$

The transformed parameters of the log-normal distributions are expressed as functions of $S_i(t_0)$ and $S_i(t_1)$:

$$\gamma_{i0} = \frac{\log(t_1)\phi^{-1}(S_i(t_0)) - \log(t_0)\phi^{-1}(S_i(t_1))}{\log(t_0) - \log(t_1)}$$

$$\gamma_{i1} = \frac{-\phi^{-1}(S_i(t_0)) + \phi^{-1}(S_i(t_1))}{\log(t_0) - \log(t_1)}$$

Therefore, rather than eliciting beliefs about the joint distribution, $\pi(\mu_1, \sigma_1, \mu_2, \sigma_2)$, we elicit beliefs about $S_i(t_0)$ and $S_i(t_1)$. To induce correlation between $S_i(t_1)$ and $S_i(t_0)$, we can elicit beliefs about the following functional parameters and assume them to be independent:[14]

$$S_1(t_0)$$

$$\delta_{11} = S_1(t_0) - S_1(t_1)$$

$$\delta_{21} = S_2(t_0) - S_1(t_0)$$

$$\delta_{22} = S_2(t_0) - S_2(t_1)$$

Beliefs about $S_1(t_0)$, $\delta_{11}$ and $\delta_{22}$ can be represented using beta distributions. Beliefs about $\delta_{21}$ can be represented using a normal distribution or as a scaled beta distribution to allow for the possibility of asymmetry in the prior distribution.

The necessary constraints to ensure plausible parameter values are:

$S_1(t_0) + \delta_{21} > 0$

$1 - S_1(t_0) - \delta_{21} > 0$

$S_1(t_0) - \delta_{11} > 0$

$S_1(t_0) + \delta_{21} - \delta_{22} > 0$

$S_1(t_1) > 0$

$S_2(t_1) > 0$

Constraints could be placed on $\gamma_{i0}$ by recognising that $S_i(1) = \phi(-\gamma_{i0})$.

Constraints on model parameters do not affect the distributions of the elicited quantities.

The expected value of a lognormal distribution is:

$$E[X] = \exp(\mu + 0.5 \times \sigma^2)$$

$$= \exp\left(-\frac{\gamma_0}{\gamma_1} + \frac{0.5}{\gamma_1^2}\right)$$

The median of a lognormal distribution is:

$$\mathrm{Med}[X] = \exp(\mu)$$

$$= \exp\left(-\frac{\gamma_0}{\gamma_1}\right)$$

*Log-logistic distribution: Joint prior distribution for the shape and scale parameters of multiple treatments*

A log-logistic distribution is characterised by a shape parameter β>0 and scale parameter θ>0. It can be derived as a mixture of a Weibull distribution with exponential mixing weights.

For a control and experimental treatment, the task is to generate a prior joint distribution for $\beta_i$ and $\theta_i$. Let $S_i(t_0)$ and $S_i(t_1)$ represent the proportion of patients out of the whole population defined by a study's inclusion/exclusion criteria who survive at times $t_0$ and $t_1$ on Treatment $i$, $i = 1,2$, where $t_1 > t_0$.

Now, the failure odds of a log-logistic model is:

$$\frac{1 - S(t)}{S(t)} = \left(\frac{t}{\theta}\right)^\beta$$

Therefore,

$$\log\left(\frac{1 - S(t)}{S(t)}\right) = \beta(\log(t) - \log(\theta))$$

$$= \alpha + \beta\log(t)$$

where

$$\alpha = -\beta\log(\theta)$$

The parameters of the log-logistic distributions are expressed as functions of $S_i(t_0)$ and $S_i(t_1)$:

$$\beta_i = \frac{[\log(1 - S_i(t_0)) - \log(S_i(t_0))] - [\log(1 - S_i(t_1)) - \log(S_i(t_1))]}{\log(t_0) - \log(t_1)}$$

$$\log(\theta_i) = \frac{\log(t_1)[\log(1 - S_i(t_0)) - \log(S_i(t_0))] - \log(t_0)[\log(1 - S_i(t_1)) - \log(S_i(t_1))]}{\beta_i(\log(t_0) - \log(t_1))}$$

Therefore, rather than eliciting beliefs about the joint distribution, $\pi(\beta_1, \theta_1, \beta_2, \theta_2)$, we elicit beliefs about $S_i(t_0)$ and $S_i(t_1)$. To induce correlation between $S_i(t_1)$ and $S_i(t_0)$, we can elicit beliefs about the following functional parameters and assume them to be independent:[14]

$$S_1(t_0)$$

$$\delta_{11} = S_1(t_0) - S_1(t_1)$$

$$\delta_{21} = S_2(t_0) - S_1(t_0)$$

$$\delta_{22} = S_2(t_0) - S_2(t_1)$$

Beliefs about $S_1(t_0)$, $\delta_{11}$ and $\delta_{22}$ can be represented using beta distributions. Beliefs about $\delta_{21}$ can be represented using a normal distribution or as a scaled beta distribution to allow for the possibility of asymmetry in the prior distribution.

The necessary constraints to ensure plausible parameter values are:

$S_1(t_0) + \delta_{21} > 0$

$1 - S_1(t_0) - \delta_{21} > 0$

$S_1(t_0) - \delta_{11} > 0$

$S_1(t_0) + \delta_{21} - \delta_{22} > 0$

$S_1(t_1) > 0$

$S_2(t_1) > 0$

Constraints could be placed on $\alpha_i$ by recognising that $S_i(1) = [1 + e^{\alpha_i}]^{-1}$

Constraints on model parameters do not affect the distributions of the elicited quantities.

The expected value of a log-logistic distribution is:

$$E[t] = \theta\Gamma\left(1 + \frac{1}{\beta}\right)\Gamma\left(1 - \frac{1}{\beta}\right)$$

The expected value does not exist if $\beta \leq 1$.

Alternatively, using the properties that:

$$\Gamma(x + 1) = x\Gamma(x)$$

$$\Gamma(1 - x) = -x\Gamma(-x)$$

$$\sin(\pi x) = \frac{\pi}{-x\Gamma(x)\Gamma(-x)}$$

The expected value of a log-logistic distribution is:

$$E[t] = \frac{\theta\pi/\beta}{\sin(\pi/\beta)}$$

The median of a log-logistic distribution is:

$$\text{Med}[t] = \theta$$

.

*Gompertz distribution: Joint prior distribution for the location and shape parameters of multiple treatments*

A Gompertz distribution is characterised by a shape parameter θ>0 and scale parameter λ>0.

For a control and experimental treatment, the task is to generate a prior joint distribution for $\theta_i$ and $\lambda_i$. Let $S_i(t_0)$ and $S_i(t_1)$ represent the proportion of patients out of the whole population defined by a study's inclusion/exclusion criteria who survive at times $t_0$ and $t_1$ on Treatment $i$, $i = 1,2$, where $t_1 > t_0$.

Let time be divided into 3 time intervals, $[x_0, x_1], (x_1, x_2], (x_2, x_3]$, where $x_0 = 0$ and $x_3 = \infty$.

Now,

$$\pi = P(x_j < T \leq x_{j+1} | T > x_j) = 1 - \frac{S(x_{j+1})}{S(x_j)}$$

When the time intervals are short, the hazard rates within each time interval can be approximated to be constant such that:

$$h(x) = -\frac{\log(1 - \pi)}{\Delta x}$$

Setting $x_1 = t_0$ and $x_2 = t_1$,

$$\lambda_i e^{\theta_i t_0} = -\frac{\log\left(\frac{S(t_0)}{S(x_0)}\right)}{t_0 - x_0}$$

$$\lambda_i e^{\theta_i t_1} = -\frac{\log\left(\frac{S(t_1)}{S(t_0)}\right)}{t_1 - t_0}$$

$$\Rightarrow \log(\lambda_i) + \theta_i t_0 = \log\left(-\frac{\log\left(\frac{S(t_0)}{S(x_0)}\right)}{(t_0 - x_0)}\right)$$

$$\Rightarrow \log(\lambda_i) + \theta_i t_1 = \log\left(-\frac{\log\left(\frac{S(t_1)}{S(t_0)}\right)}{(t_1 - t_0)}\right)$$

The parameters of the Gompertz distributions are expressed as functions of $S_i(t_0)$ and $S_i(t_1)$:

$$\theta_i = \frac{\log\left(-\frac{\log\left(\frac{S(t_1)}{S(t_0)}\right)}{(t_1 - t_0)}\right) - \log\left(-\frac{\log\left(\frac{S(t_0)}{S(x_0)}\right)}{(t_0 - x_0)}\right)}{t_1 - t_0}$$

$$\log(\lambda_i) = \frac{t_1 \log\left(-\frac{\log\left(\frac{S(t_0)}{S(x_0)}\right)}{(t_0 - x_0)}\right) - t_0 \log\left(-\frac{\log\left(\frac{S(t_1)}{S(t_0)}\right)}{(t_1 - t_0)}\right)}{t_1 - t_0}$$

Therefore, rather than eliciting beliefs about the joint distribution, $\pi(\theta_1, \lambda_1, \theta_2, \lambda_2)$, we elicit beliefs about $S_i(t_0)$ and $S_i(t_1)$. To induce correlation between $S_i(t_1)$ and $S_i(t_0)$, we can elicit beliefs about the following functional parameters and assume them to be independent:[14]

$$S_1(t_0)$$

$$\delta_{11} = S_1(t_0) - S_1(t_1)$$

$$\delta_{21} = S_2(t_0) - S_1(t_0)$$

$$\delta_{22} = S_2(t_0) - S_2(t_1)$$

Beliefs about $S_1(t_0)$, $\delta_{11}$ and $\delta_{22}$ can be represented using beta distributions. Beliefs about $\delta_{21}$ can be represented using a normal distribution or as a scaled beta distribution to allow for the possibility of asymmetry in the prior distribution.

The necessary constraints to ensure plausible parameter values are:

$S_1(t_0) + \delta_{21} > 0$

$1 - S_1(t_0) - \delta_{21} > 0$

$S_1(t_0) - \delta_{11} > 0$

$S_1(t_0) + \delta_{21} - \delta_{22} > 0$

$S_1(t_1) > 0$

$S_2(t_1) > 0$

The prior survival functions approximate the elicited quantities as a consequence of approximating the true hazards at times $t_0$ and $t_1$.

The expected value of a Gompertz distribution is:

$$E[t] = \frac{1}{\theta} e^{(\lambda/\theta)} \Gamma\left(0, \lambda/\theta\right)$$

$\Gamma\left(0, \lambda/\theta\right)$ is the incomplete exponential integral, a special case of the upper incomplete gamma function defined as $\Gamma(a, x) = \int_x^\infty t^{a-1} e^{-t} dt$.

The median of a Gompertz distribution is:

$$\text{Med}[t] = \frac{1}{\theta} \log\left[1 - \left(\frac{\theta}{\lambda}\right) \log\left(\frac{1}{2}\right)\right]$$

## Computing survival probabilities and mean survival

Let $\psi_t(\theta_m)$ represent the proportion of participants surviving after $t$ months as a function of the model parameters. The Bayesian estimate of the survival function is computed as the expected value of its posterior distribution:

$$S_m(t) = E[\psi_t(\theta_m)].$$

Similarly, let $\varphi(\theta_m)$ represent the population mean survival as a function of the model parameters. The Bayesian estimate is computed as the expected value of its posterior distribution:

$$E[\varphi(\theta_m)].$$

## Model probabilities

### *Non-informative prior distributions*

There is no unique way to represent so-called non-informative prior distributions for model probabilities. Options include:

- Let $f(m) = 1/M$.

- If $f(1)$ represents the prior belief in model 1, then $f(m) = (1 - f(1))/(M - 1)$, $m = 2, \ldots, M$.

- If models are nested such that Model 1 $\subset$ Model 2 $\subset \ldots \subset$ Model M, then smaller prior probabilities are placed on models of higher dimension:

    - Jeffrey's improper prior $f(m) = 1/(m + 1)$

    - Suppose there are $n_s$ models of dimension $\dim(m_s)$, $s = 1, \ldots, S$. Then, let the prior probability be $f(m_s)$ for models of dimension $\dim(m_s)$[16]

1. Give each model of dimension $\dim(m_s)$ equal probability $f(m_s) = 1/S$ so that $f(m) = f(m_s)/n_s$.

2. Give each model of dimension $\dim(m_s)$ probability $f(m_s) = m_s^{-1}/\sum_{s=1}^{S} s^{-1}$ so that $f(m) = f(m_s)/n_s$.

*Empirical prior distributions*

Garthwaite et al. proposed an empirical Bayes approach based on formulating correlation matrices between models.[17] Three weighting schemes were proposed based on the correlation matrix: minimum variance, capped eigenvalue and cos-square. The cos-square weighting scheme satisfies the strong dilution property and has weights that are always non-negative.

*Computing Bayesian model evidence using Monte Carlo simulation*

Step 1:

Generate N random draws from the prior joint distribution of model parameters, $\theta$.

Step 2:

For each random draw $(i = 1, \dots, N)$ from the prior joint distribution of model parameters, compute the likelihood $(j = 1, \dots, J)$.

$$\text{lik}_i = \prod_{j}^{J} [h(t_j|\theta_i)]^{\delta_j} S(t_j|\theta_i)$$

Step 3:

Compute the mean of the likelihoods.

$$BME = f_m(\mathcal{D}) = \frac{\sum_{i=1}^{N} \text{lik}_i}{N}$$

*Assessing Bayesian model evidence convergence*

Compute the cumulative average of the likelihoods and plot against $N$.

$$\text{BME}_N = \frac{\sum_{i=1}^{N} \text{lik}_i}{N}$$

$N$ is sufficient when $\text{BME}_N$ stabilises.

*Computing Hellinger's distance using Monte Carlo simulation*

Define

$$f_{D^H}(m) \propto a_m$$

where

$$a_m = \sum_{m'} D^H(m, m')$$

the sum of the distances from model $m$ to every other model.

The aim is to compute:

$$D^H(m, m') = \int \left[ f_m^{1/2}(y) - f_{m'}^{1/2}(y) \right]^2 dy$$

where the integration is over the sample space of the data for models $m$ and $m'$ and

$$f.(y) = \int f.(y|\theta.) f.(\theta.) d\theta.$$

Now, using Monte Carlo simulation:

$$\int \left[ f_m^{1/2}(y) - f_{m'}^{1/2}(y) \right]^2 dy \approx \frac{1}{J} \sum_{j=1}^{J} \left[ f_m^{1/2}(y_j) - f_{m'}^{1/2}(y_j) \right]^2$$

where $y_1, \ldots, y_J$ are independent and identically distributed uniform random variables.

Similarly,

$$f^{1/2}(y_j) \approx \frac{1}{N}\sum_{i=1}^{N} f^{1/2}(y_j|\theta_{\cdot i})$$

where $\theta_{\cdot i}$ are independent and identically distributed $f(\theta_{\cdot})$.

Step 1:

Generate a uniform distributed random survival time, $y_j$, over the sample space of the data, say $U[0,100]$.

Step 2:

Generate random draws, $\theta_{mi}$ and $\theta_{m'i}$, from the prior joint distributions of parameters from model $m$ and $m'$, respectively.

Step 3:

Compute $f(y_j|\theta_{\cdot i})$ for the $y_j$ generated at Step 1 with respect to parameter values $\theta_{mi}$ and $\theta_{m'i}$ generated at Step 2 for models m and m', respectively.

Step 4:

Compute the square root of $f(y_j|\theta_{\cdot i})$ generated at Step 3 for models m and m', respectively

Step 5:

Repeat Steps 2-4 for a large number draws, $i = 1, \ldots, N$, from the prior joint distributions of parameters from model m and m'

Step 6:

Compute the average of the values generated at Step 5 with respect to each model.

Step 7:

Compute the difference of the averages of the square roots computed at Step 6 and square the result.

Step 8:

Repeat steps 1-7 for a large number of random sample times, $j = 1, \ldots, J$.

Step 9:

Compute the average of the values computed at Step 8.

## Appendix S2: German breast cancer study

Table S2.1 summarises the number of recurrence-free events and the ages of the patients at the start of treatment.

Table S2.1: Number of recurrence free events

|  | **No Hormonal Treatment (N=440)** | **Hormonal Treatment (N=246)** |
|---|---|---|
| Recurrence-free events | 205(46.6%) | 94 (38.2%) |
| Age yrs: Mean | 51 | 57 |
| Minimum | 21 | 32 |

Exponential Distribution

The prior and posterior distributions of model parameters, survival functions and Bayesian model evidence are presented in Figure S2.1.

Figure S2.1: Exponential distribution: Distribution of model parameters, survival functions and Bayesian model evidence]

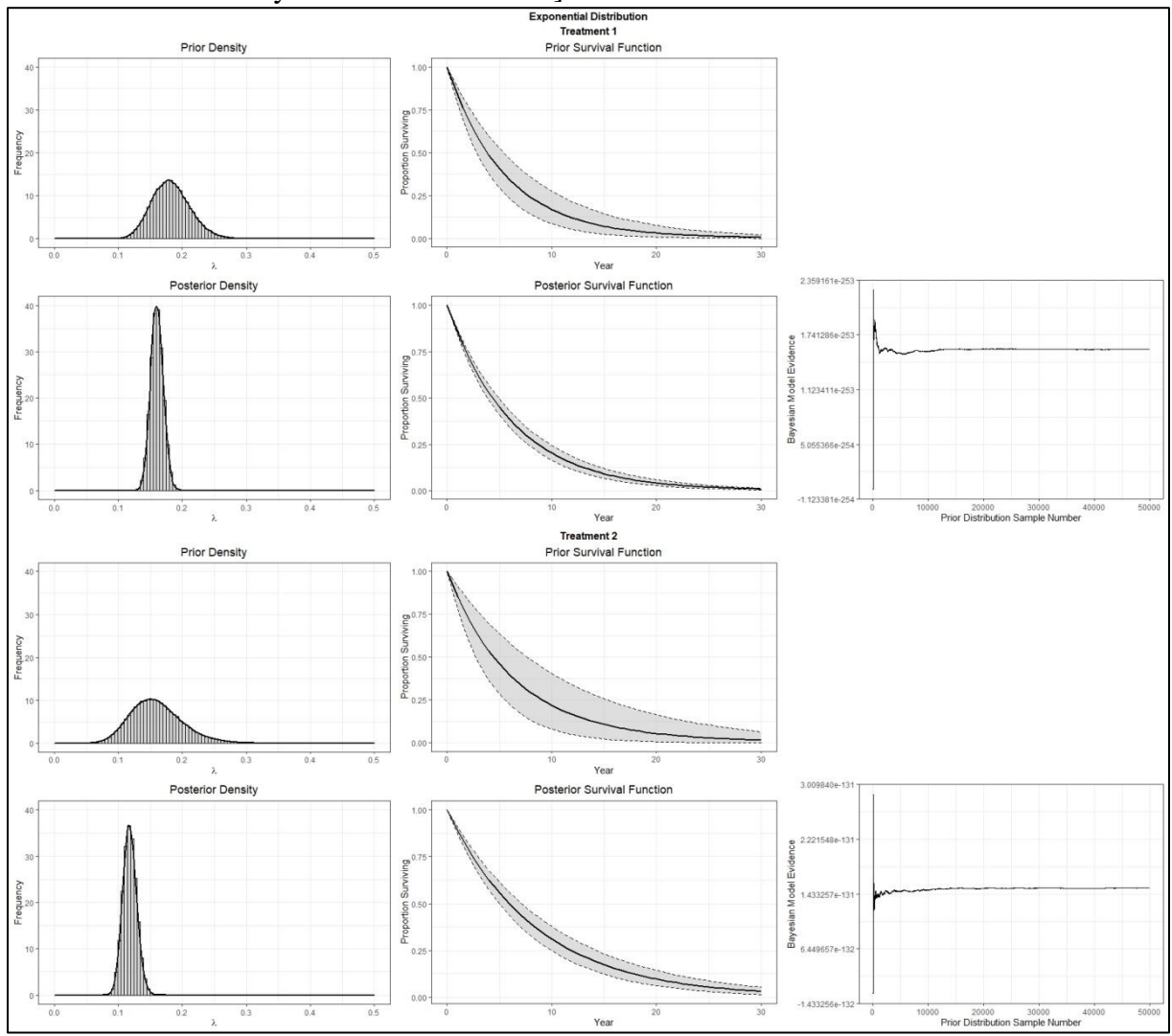

Weibull Distribution

When a sampled value from the prior distribution for $\delta_{ii}$ is small between times $t_0$ and $t_1$, $v_i$ tends to zero and a proportion of patients tend to immortality. Similarly, when a sampled value from the prior distribution for $\delta_{ii}$ is large between times $t_0$ and $t_1$, $v_i$ tends to infinity and a proportion of patients die quickly. According to Selvin, the usual values of the shape parameter of a Weibull distribution used to model human survival data are $0.3 \leq v_i \leq 3.5$.[18] Using this constraint in the illustrative example generated several prior pairs of $(v_i, \lambda_i)$ corresponding to prior mean values that were greater than 50 years, while the average age of patients in the study was 53 years. To avoid implausibly high proportions of patients living beyond 100 years, we added the constraint that the population means should be no more than 50 years. A population mean survival of 50 years might still be implausible and, with clinical input, a stricter constraint could be used.

Lognormal Distribution

The illustrative example generated several prior pairs of $(\mu_i, \sigma_i)$ corresponding to prior mean values that were greater than 50 years. To avoid implausibly high proportions of patients living beyond 100 years, we added the constraint that the population means should be no more than 50 years. A population mean of 50 years might be implausible and, with clinical input, a stricter constraint could be used.

The prior and posterior distributions of model parameters, survival functions and Bayesian model evidence are presented in Figure S2.2.

Figure S2.2: Lognormal distribution: Joint distributions of model parameters, survival functions and Bayesian model evidence

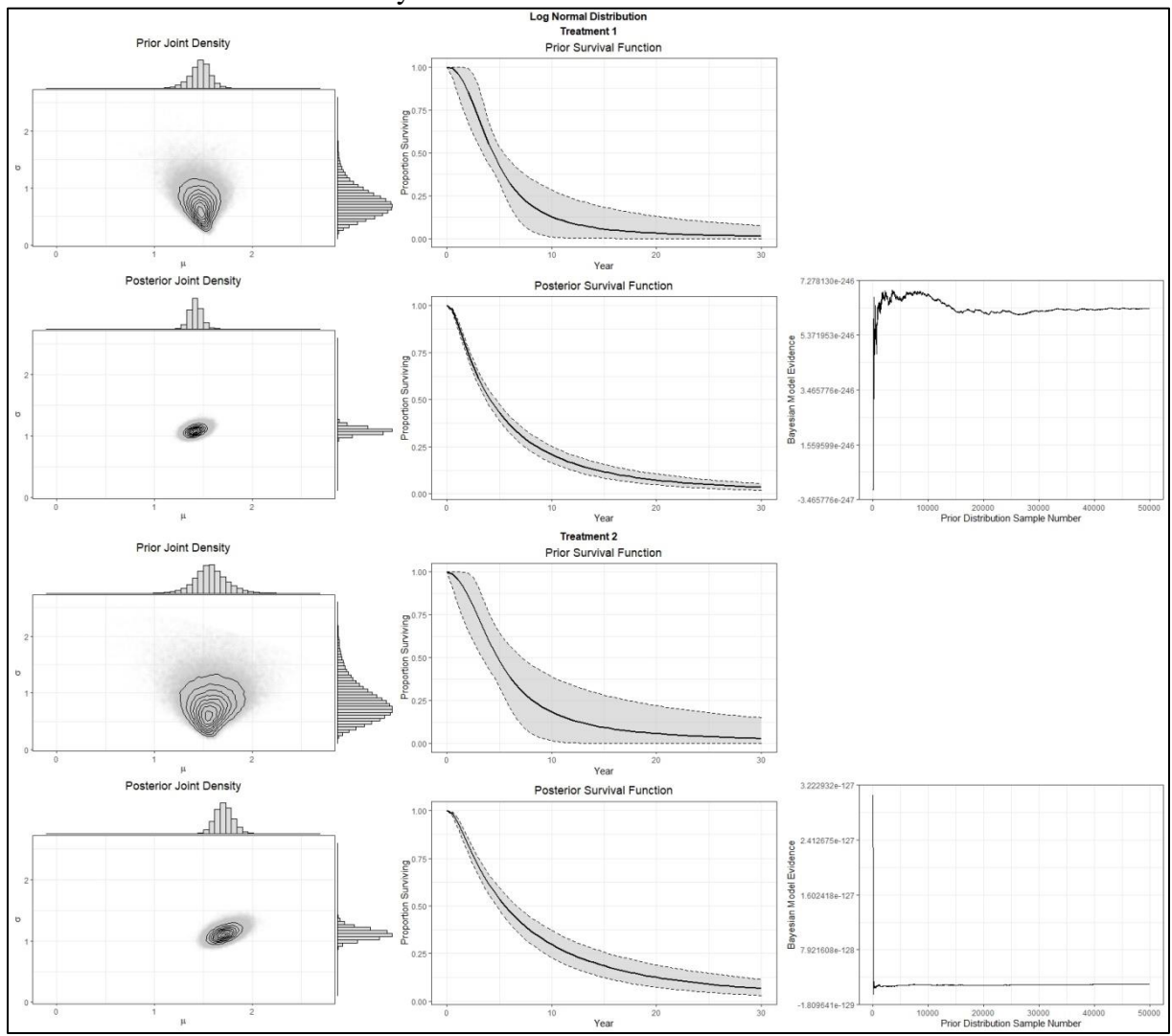

Log-logistic Distribution

The illustrative example generated several prior pairs of $(\beta_i, \theta_i)$ corresponding to prior mean values that were greater than 50 years. To avoid implausibly high proportions of patients living beyond 100 years, we added the constraint that the population means should be no more than 50 years. A population mean of 50 years might be implausible and, with clinical input, a stricter constraint could be used.

The prior and posterior distributions of model parameters, survival functions and Bayesian model evidence are presented in Figure S2.3.

Figure S2.3:   Log-logistic distribution: Joint distributions of model parameters, survival functions and Bayesian model evidence

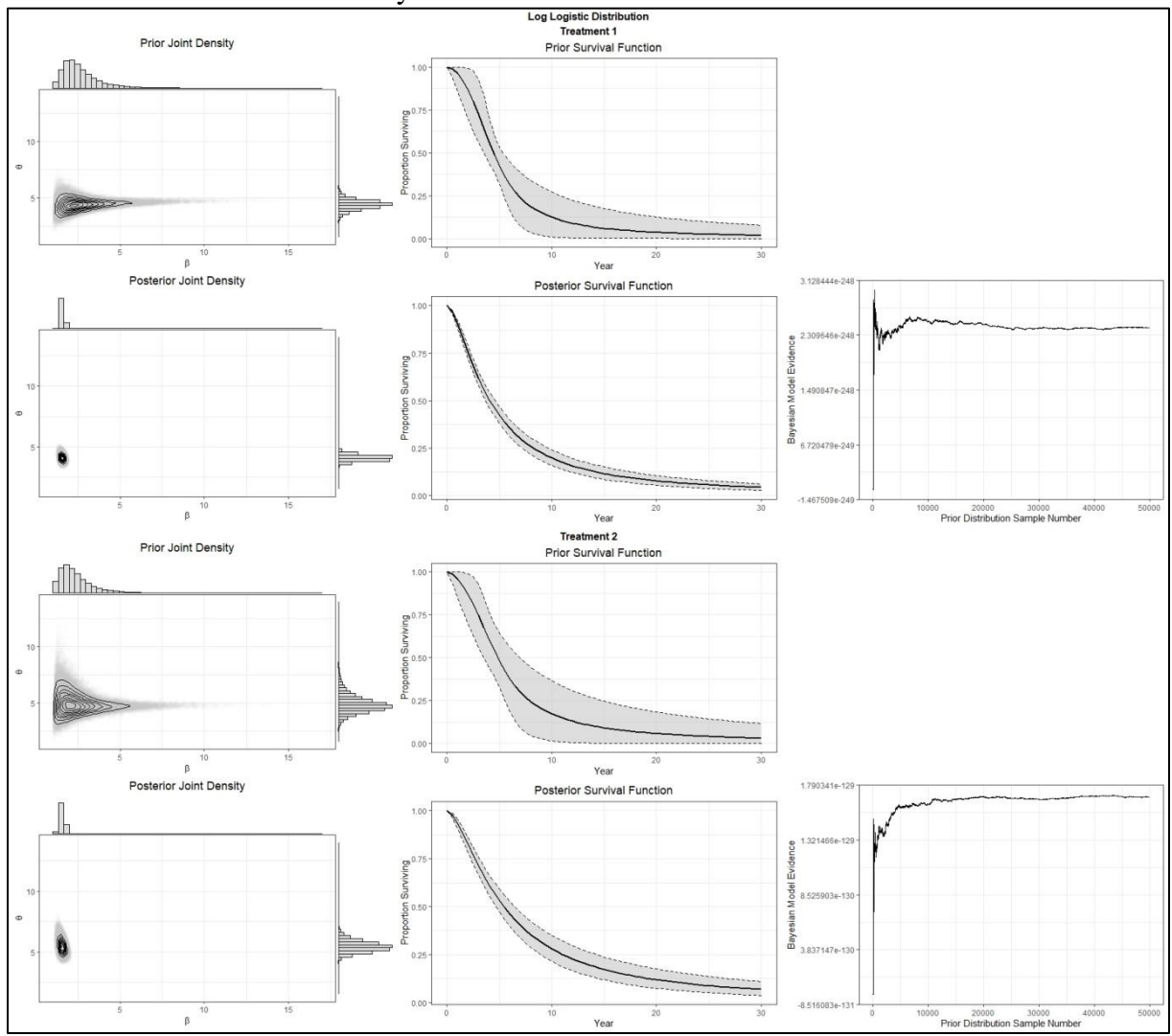

Gompertz Distribution

Only approximately 23% of samples from the prior joint distribution satisfied the constraint that $\theta$ must be greater than $\lambda$ for a plausible survival model. This suggests *a priori* that a Gompertz distribution is unlikely to be a model that is consistent with prior beliefs about the proportions of patients surviving.

The prior and posterior distributions of model parameters, survival functions and Bayesian model evidence are presented in Figure S2.4.

Figure S2.4: Gompertz distribution: Joint distributions of model parameters, survival functions and Bayesian model evidence

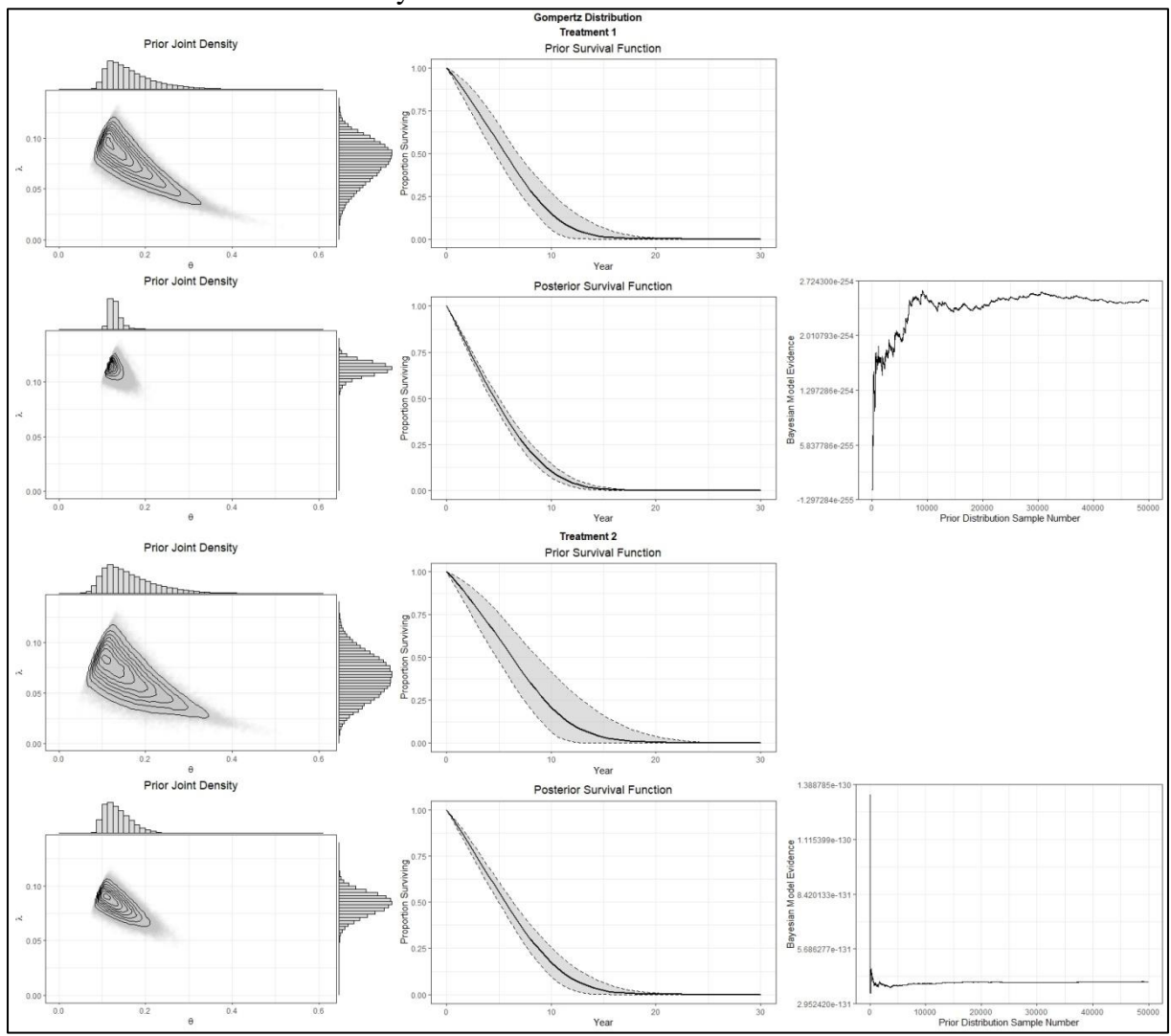

Gompertz Distribution

References

1. Schoniger A, Wohling T, Samaniego L, Nowak W. Model selection on solid ground: Rigorous comparison of nine ways to evaluate Bayesian model evidence. *Water Resources Research* 2014;50:9484-953.
2. National Institute for Health and Care Excellence. Guide to the methods of technology appraisal; 2013.
3. O'Hagan A, Buck C, Daneshkhah A, Eiser J, Garthwaite P, Jenkinson D*, et al.* Uncertain judgements: Eliciting experts' probabilities: John Wiley & Sons Ltd,; 2006.
4. European Food Safety Authority. Guidance on expert knowledge elicitation in food and feed safety risk assessment. *EFSA Journal* 2014;12.
5. Dallow N, Best N, Montague T. Better decision making in drug development through adoption of formal prior elicitation. *Pharmaceutical Statistics* 2018;17:301-16.
6. O'Hagan A. Expert knowledge elicitation: subjective but scientific. *The American Statistician* 2019;73:69-81.
7. Williams C, Wilson K, Wilson N. A comparison of prior elicitation aggregation using the classical method and SHELF. *Journal of the Royal Statistical Society Series A* 2021; DOI: 10.1111/rssa.12691.
8. Gore S. Biostatistics and the Medical Research Council *Medical research Council News* 1987;35:19-20.
9. Morris D, Oakley J, Crowe J. A web-based too for eliciting probability distributions from experts. *Envoronmental Modelling &Software* 2014;52:1-4.
10. Oakley J, O'Hagan A. SHELF: the Sheffield Elicitation Framework (version 3.0). School of Mathematics and Statistics, University of Sheffield, UK. In; 2016.
11. Cox C. The generalised F distribution: An umbrella for parametric survival analysis. *Statistics in Medicine* 2008;27:4301-12.
12. Missov T, Lenart A, Nemeth L, Canudas-Romo V, Vaupel J. The Gompertz force of mortality in terms of the modal age at death. *Demographic Research* 2015;32.
13. Lunn D, Thomas A, Best N, Spiegelhalter D. WinBUGS - a Bayesian modelling framework: concepts, structure, and extensibility. *Statistics and Computing* 2000;10:325-37.
14. Ren S, Oakley J. Assurance calculations for planning clinical trials with time-to-event outcomes. *Statistics in Medicine* 2014;33:31-45.
15. Plummer M. A program for analysis of Bayesian graphical models using Gibbs sampling. . In. Proceedings of the 3rd International Workshop on Distributed Statistical Computing (DSC 2003), March 20-22, Vienna, Austria ISSN 1609-395X; 2003.
16. Berger J, Pericchi L. The intrinsic Bayes factor for model selection and prediction. *Journal of the American Statistical Association* 1996;91:109-22.
17. Garthwaite P, Mubwandarikwa E. Selection of weights for weighted model averaging. *Australian & New Zealand Journal of Statistics* 2010;52:363-82.
18. Selvin S. Survival analysis for epidemiologic and medical research: A practical guide: Cambridge University Press; 2008.